\RequirePackage{fix-cm}
\documentclass[twocolumn,epjc3]{svjour3}  
\smartqed  
\RequirePackage{graphicx}
\usepackage{amsmath}
\usepackage{amssymb} 
\journalname{Eur. Phys. J. C}
\begin{document}

\title{Cosmological singularities in $f(T,\phi) $ gravity  }


\author{Oem Trivedi \thanksref{e1,addr1}
        \and
        Maxim Khlopov\thanksref{e2,addr2,addr3,addr4} \and Jackson Levi Said \thanksref{e3,addr5,addr6} \and Rafael Nunes \thanksref{e4,addr7,addr8} 
}

\thankstext{e1}{e-mail: oem.t@ahduni.edu.in}
\thankstext{e2}{e-mail: khlopov@apc.in2p3.fr }
\thankstext{e3}{email: jackson.said@um.edu.mt}
\thankstext{e4}{email: rafadcnunes@gmail.com}

\institute{School of Arts and Sciences, Ahmedabad Univeristy,Ahmedabad 380009, India  \label{addr1}
           \and
           Institute of Physics, Southern Federal University, Stachki 194 Rostov on Don 344090, Russia  \label{addr2}
           \and
           Virtual Institute of Astroparticle physics, Paris 75018, France \label{addr3}
           \and
           National Research Nuclear University ”MEPHI”, Moscow 115409, Russia \label{addr4}
           \and 
           Institute of Space Sciences and Astronomy, University of Malta \label{addr5}
           \and 
           Department of Physics, University of Malta \label{addr6}
           \and 
           Instituto de F\'isica, Universidade Federal do Rio Grande do Sul, 91501-970 Porto Alegre RS, Brazil \label{addr7}
           \and 
           Divis\~ao de Astrof\'isica, Instituto Nacional de Pesquisas Espaciais, Avenida dos Astronautas 1758, S\~ao Jos\'e dos Campos, 12227-010, SP, Brazil \label{addr8} 
}

\date{Received: date / Accepted: date}

\maketitle

\begin{abstract}
    The pursuit of understanding the mysteries surrounding dark energy has sparked significant interest within the field of cosmology. While conventional approaches, such as the cosmological constant, have been extensively explored, alternative theories incorporating scalar field-based models and modified gravity have emerged as intriguing avenues. Among these, teleparallel theories of gravity, specifically the $f(T,\phi)$ formulation, have gained prominence as a means to comprehend dark energy within the framework of teleparallelism. In this study, we investigate two well-studied models of teleparallel dark energy and examine the presence of cosmological singularities within these scenarios. Using the Goriely-Hyde procedure, we examine the dynamical systems governing the cosmological equations of these models. Our analysis reveals that both models exhibit Type IV singularities, but only for a limited range of initial conditions. These results could indicate a potential edge for teleparallel cosmological models over their other modified gravity counterparts, as the models we examine seem to be only allowing for weak singularities that too under non general conditions. 
\end{abstract}

\section{Introduction}

Observations of the late-time acceleration of the Universe came as a surprise to the cosmological community \cite{SupernovaSearchTeam:1998fmf}. Since then, extensive efforts have been dedicated to explaining this expansion. Standard approaches, such as the Cosmological constant \cite{SupernovaSearchTeam:1998fmf,Weinberg:1988cp,Lombriser:2019jia,Copeland:2006wr,Padmanabhan:2002ji}, as well as more exotic scenarios like Modified gravity theories \cite{Capozziello:2011et,Nojiri:2010wj,Nojiri:2017ncd}, and recent proposals for the direct detection of dark energy \cite{Zhang:2021ygh}, have been pursued. One fascinating avenue for understanding dark energy is through Quintessence, where a scalar field drives the late-time cosmic acceleration of the universe \cite{Zlatev:1998tr,Tsujikawa:2013fta,Faraoni:2000wk,Gasperini:2001pc,Capozziello:2003tk,Capozziello:2002rd,Carroll:1998zi,Caldwell:2005tm,Han:2018yrk,Astashenok:2012kb,Shahalam:2015sja}. Quintessence is particularly interesting as it represents the simplest scalar field dark energy scenario that avoids issues like ghosts or Laplacian instabilities. In quintessence models, the acceleration of the universe is driven by a slowly varying scalar field with a potential $V(\phi)$, similar to the mechanism of slow-roll inflation. However, in this case, contributions from non-relativistic matter, such as baryons and dark matter, cannot be neglected.
\\
\\
It is worth noting that simple models of Quintessence have been shown to be in conflict with the current H0 tension \cite{Colgain:2019joh,Banerjee:2020xcn,DiValentino:2021izs}, suggesting that simple Quintessence models may perform worse than $\Lambda$-CDM models in light of the current H0 data \cite{riess2021comprehensive}.This leads one to consider other more exotic possibilities for scalar field dark energy models and one such possibility is to consider models in Teleparallel gravity.Teleparallel gravity, a theory based on torsion, provides an alternative description of gravity \cite{1985FoPh...15..365I,pellegrini1963tetrad,Hayashi:1979qx,Maluf:1994ji,de2000teleparallel,Arcos:2004tzt,pereira2014teleparallelism,Pereira:2019woq,Leon:2012mt}, where gravitation is mediated by torsion. In this approach, the Lagrangian density of Teleparallel Equivalent of General Relativity (TEGR) is proportional to the torsion scalar $T$. In TEGR, the tetrad field and spin connection pair replace the metric tensor and Levi-Civita connection, respectively, while the teleparallel connection replaces the usual connection \cite{pereira2014teleparallelism,Arcos:2004tzt}. Consequently, at the level of the dynamical equations, curvature-based gravitational theories are equivalent to tensor-based theories \cite{Arcos:2004tzt,Aldrovandi:2013wha}. By introducing an arbitrary function $f(T)$ in place of the torsion scalar $T$, a generalization of TEGR known as $f(T)$ gravity is obtained \cite{Ferraro:2008ey,Bengochea:2008gz,Linder:2010py,Basilakos:2013rua,Finch:2018gkh,Bahamonde:2017ize,Basilakos:2018arq}, leading to new cosmological models. In this framework, the tetrad fields, which form the orthogonal basis for the tangent space, serve as the dynamical variables of teleparallel gravity. The torsion tensor is constructed using the first derivative of the tetrad product. The field equations are derived by varying the action with respect to the tetrad fields, while the spin connection preserves the local Lorentz invariance and contributes to the equations of motion. For further exploration of $f(T)$ gravity, refer to \cite{Wu:2010xk,Dent:2010nbw,Farrugia:2016qqe,Cai:2019bdh,Briffa:2020qli,LeviSaid:2021yat,Duchaniya:2022rqu}. The investigation of scalar-torsion theories with non-minimal coupling between the torsion scalar and a scalar field was carried out in the context of dark energy \cite{Geng:2011aj,Geng:2011ka}, including studies with arbitrary non-minimal coupling functions and tachyon terms for the scalar field \cite{Otalora:2013dsa,Otalora:2013tba}. Another extension of $f(T)$ gravity is the generalized scalar-torsion $f(T,\phi)$ gravity, where $\phi$ represents the canonical scalar field, and the gravitational action incorporates a non-minimal coupling between the scalar field and torsion scalar \cite{Xu:2012jf}. Additionally, within the covariant teleparallel framework, a new class of theories has been proposed, where the action depends on the scalar field and an arbitrary function of the torsion scalar \cite{Hohmann:2018rwf}.
\\
\\
Recently, a significant amount of research has been dedicated to exploring the various types of cosmological singularities that may occur in the present and distant future of the Universe \cite{Nojiri:2004ip,Nojiri:2005sr,Nojiri:2005sx,Bamba:2008ut,Bamba:2010wfw,Nojiri:2008fk,odintsov2022did,deHaro:2023lbq,Nojiri:2006ww,Bamba:2012vg,Capozziello:2009hc,Nojiri:2004pf,Elizalde:2004mq,Abdalla:2004sw}. However, it is often challenging to classify and study cosmological singularities in highly unconventional cosmologies influenced by considerations of quantum gravity or phenomenology. Traditional methods may not be applicable in these cases. Therefore, alternative approaches are necessary to identify cosmological singularities in exotic cosmologies. In this context, the Goriely-Hyde procedure, a particular method in dynamical systems, can be extremely useful \cite{goriely2000necessary}. Understanding the singularity structure of dynamical systems is an intriguing aspect, especially when these systems describe significant physical phenomena. Although various approaches have been proposed to investigate the singularity structure of autonomous dynamical systems, the Goriely-Hyde procedure has proven particularly valuable for cosmological studies due to the abundance of interesting dynamical systems in cosmology \cite{bahamonde2018dynamical}. Previous applications of the Goriely-Hyde method have explored finite and non-finite time singularities in specific quintessence models \cite{odintsov2018dynamical,odintsov2019finite,Trivedi:2022ipa}. However, a comprehensive analysis of cosmological singularities teleparallel  models of dark energy using this approach is still lacking, and our work aims to address this gap.The study of the cosmological dynamics and stability of $f(t,\phi)$ dark energy was conducted in \cite{Gonzalez-Espinoza:2020jss}, while an analysis of scalar perturbations was performed in \cite{Gonzalez-Espinoza:2021mwr}. A recent full dynamical systems analysis of $f(t,\phi) $ dark energy for two particular models was done in \cite{Duchaniya:2022fmc} and we intend to use the dynamical systems approach developed there to pursue our singularity analysis. In Section II, we provide a concise overview of Teleparallel gravity method while in section III we provide a brief review of the Goriely-Hyde method. In Section IV we apply the Goriely-Hyde method to two particular models and demonstrate the diverse characteristics of singularities in $f(T,\phi)$ models, including both finite and infinite-time occurrences which can occur for both $f(t,\phi)$ models considered in \cite{Duchaniya:2022fmc} . Subsequently, in Section V, we consider two well-motivated ansatz for the Hubble parameter and classify the types of cosmological singularities (Types I-IV) that can arise within these regimes. Finally, we conclude our work in Section VI.
\section{Teleparallel gravity}
General relativity (GR) can account for most observed phenomena with appropriate modifications considered in the matter sector. In this context, the widely accepted concordance model combines $\Lambda$-CDM cosmology with inflation. However, the enigmatic nature of certain particle species remains a puzzle despite significant progress in physics beyond the standard model of particle physics. It is also plausible that the standard model of particle physics might not require substantial restructuring to address these observational challenges. Instead, it could be the gravitational sector that requires further examination. This could involve extensions of GR or modifications beyond GR as alternatives to its original formulation. The scientific literature has witnessed numerous proposals for new theories of gravity, motivated by various phenomena, theoretical approaches, or even quantum physics. One intriguing possibility that has garnered increasing attention in recent decades is teleparallel gravity, where torsion replaces curvature as the mechanism responsible for generating gravitational fields. This theory replaces the traditional Levi-Civita connection, which is curvature-based, with a teleparallel connection based on torsion. Numerous publications on this topic have emerged in the literature. Among the theories arising from torsion-based approaches to gravity is the teleparallel equivalent of general relativity (TEGR), which is dynamically equivalent to GR and thus indistinguishable from it through classical experiments.
\\
In teleparallel gravity (TG), one typically assumes an action of the form:
\begin{equation}\label{eq:matgravaction}
    \mathcal{S}_{\text{TG}} := \mathcal{S}_{\text{g}}[e, \omega] + \mathcal{S}_{\text{m}}[e, \chi]\,,
\end{equation}
Here, the gravitational part \(\mathcal{S}_{\text{g}}\) of the action depends on the tetrad \(e^A{}_{\mu}\) and the spin connection \(\omega^A{}_{B\mu}\), while the matter part depends on the tetrad \(e^A{}_{\mu}\) and arbitrary matter fields \(\chi^I\), but not on the spin connection\cite{Koivisto:2018aip,BeltranJimenez:2020sih}. This is because we assume that the hypermomentum vanishes, thereby preventing this coupling. Introducing a dependence on spin would effectively introduce a second matter tensor, resulting from the variation of the matter Lagrangian with respect to the spin connection. The variation of the matter part of the action, after integration by parts to eliminate derivatives acting on field variations, can be expressed as follows:
\begin{equation}\label{eq:matactvar}
    \delta\mathcal{S}_{\text{m}} = \int d^4 x e (\Theta_A{}^{\mu}\delta e^A{}_{\mu} + \Omega_I\delta\chi^I)\
\end{equation}
Here, \(\Omega_I = 0\) represents the matter field equations, and \(\Theta_A{}^{\mu}\) denotes the energy-momentum tensor. The corresponding variation of the gravitational action takes the form:
\begin{equation}\label{eq:gravactvar}
    \delta\mathcal{S}_{\text{g}} = -\int d^4x e(W_A{}^{\mu}\delta e^A{}_{\mu} + Y_A{}^{B\mu}\delta\omega^A{}_{B\mu})\ 
\end{equation}
The tensors \(W_A{}^{\mu}\) and \(Y_A{}^{B\mu}\) arise from the variation and integration by parts, with their specific form depending on the particular theory under consideration. The explicit expression for \(W_A{}^{\mu}\) can be found for several theories in \cite{Bahamonde:2021gfp}. For brevity, we omit \(Y_A{}^{B\mu}\) here, as it turns out to be redundant in deriving the field equations, which can be entirely determined from \(W_A{}^{\mu}\) alone. Furthermore, by varying with respect to the tetrad, one can derive the field equations:
\begin{equation}\label{eq:gentetradfieldlor}
W_A{}^{\mu} = \Theta_A{}^{\mu}\,.
\end{equation}
An alternative representation of the field equations, more commonly used, is obtained by transforming the first index into a spacetime index with the tetrad while lowering the second index:
\begin{equation}
W_{\mu\nu} = e^A{}_{\mu}g_{\rho\nu}W_A{}^{\rho}\,, \quad
\Theta_{\mu\nu} = e^A{}_{\mu}g_{\rho\nu}\Theta_A{}^{\rho}\,,
\end{equation}
This yields the field equations in the form:
\begin{equation}\label{eq:gentetradfield}
W_{\mu\nu} = \Theta_{\mu\nu}\,.
\end{equation}
However, deriving the field equations for the spin connection is more complex, as it must satisfy the conditions of being flat, \(R^{\alpha}{}_{\beta\mu\nu} = 0\), and metric-compatible, \(\nabla_{\alpha}g_{\mu\nu} = 0\), by definition. Various approaches exist to maintain these properties during the variation procedure \cite{Golovnev:2017dox,Hohmann:2021fpr}.
\\
\\
For considering cosmological scenarios in teleparallel theories, it is helpful to consider an FLRW metric of the form \cite{Bahamonde:2021gfp} \begin{equation} \label{eq:frwN}
    d s^2 = N(t)^2 d t^2-a(t)^2\Big[\frac{d r^2}{1-kr^2}+r^2(d \vartheta^2+\sin^2\vartheta d \varphi^2)\Big]\,,
\end{equation}
where $N(t)$ and $a(t)$ represent the lapse function and scale factor respectively. In the case of flat universes ($k=0$) one can write \begin{equation}\label{FLRW_metric}
    \mathrm{d}s^2 = N(t)^2\mathrm{d}t^2 - a^2(t) \left(\mathrm{d}x^2+\mathrm{d}y^2+\mathrm{d}z^2\right)\,,
\end{equation}
and this results in the diagonal tetrad 
\begin{equation}\label{FLRW_tetrad}
    {e}_{{\mu}}^{A}=\text{diag}\left(N(t),\,a(t),\,a(t),\,a(t)\right)\,,
\end{equation}
which turns out to be in the Weitzenbock gauge for the extensions to TEGR. An important remark here is that the above tetrad (with vanishing spin connection) is the only one that has the property that both the tetrad and the teleparallel connection obey cosmological symmetries for flat FLRW. One can also relax the condition that the teleparallel connection enjoys the symmetries of cosmology, but then, the corresponding cosmological equations would not respect the symmetries of cosmology. If we use the diagonal tetrad~\eqref{FLRW_tetrad} in Cartesian coordinates one can, for example, find the modified FLRW equations for $f(T)$ gravity as
\begin{subequations} \label{ftfriedmann}
	\begin{align}
	-6H^2f_T- \frac{1}{2}f &= \kappa^2\rho\label{Friedmann_1B}\,, \\[0.5ex]
	-2f_T(3H^2+\dot{H}) - 2H\dot{f}_T - \frac{1}{2}f &= -\kappa^2 p\label{Friedmann_2B}\,,
	\end{align}
\end{subequations}
where dots are derivatives with respect to time, so that $\dot{f}_T=f_{TT}\dot{T}$. One can further obtain modified FLRW equations for other teleparallel appraoches, like $f(T,B)$ gravity being \begin{subequations} \label{ftbfriedmann}
\begin{align}
	3 H\dot{f}_B -3H^2 ( 3f_B+2f_T)-3 f_B\dot{H}-\frac{1}{2} f(T,B)&=\kappa ^2\rho\,,\label{eq:fTBFRW1}\\[0.5ex]
	-(3H^2+\dot{H})(2f_T+3f_B)-2 H\dot{f}_T+\ddot{f}_B-\frac{1}{2} f(T,B)&=-\kappa ^2 p\,.\label{eq:fTBFRW2}
\end{align}
\end{subequations}
Or for the Teleparallel Gauss-Bonet models being (fixing the gauge such that $N=1$) \begin{subequations} \label{gaussbonetfriedmann}
\begin{multline}
-6 f_T H^2 - 12 H^3 \dot{f}_{T_G} + 12f_{T_G} H^2 \left(\dot{H}+H^2\right) \\
- \frac{1}{2} f(T,T_G) = \kappa^2\rho  \label{eq:FRWfTG1}
\end{multline}
\begin{multline}
-2 H\dot{f}_T - 2 f_T \left(\dot{H}+3 H^2\right) + 12f_{T_G} H^2 \left(\dot{H}+H^2\right) \\
- 8 H \dot{f}_{T_G} \left(\dot{H}+H^2\right) -4 H^2 \ddot{f}_{T_G} - \frac{1}{2} f(T,T_G) = -\kappa^2p \\
  \label{eq:FRWfTG2}
\end{multline}
\end{subequations}
where $f_{T_G}=\frac{\partial f}{\partial T_G} T_G$. It is worth mentioning that $B_G=0$ in flat FLRW and then the standard Gauss-Bonnet term is just ${\mathcal{G}}=T_G$. 
\\
\\
While there are a multitude of approaches of dealing with such exotic cosmological systems like reconstruction methods or Noether-symmetries approaches, dynamical systems methods are also an efficient way to understand the dynamics of the models. Dynamical systems allow for the extraction of key features of the cosmology without solving the evolution equations directly (in an exact form). Thus, it then becomes possible to describe the overall nature of the gravitational theory and henceforth determine whether the model can generate a viable cosmological evolution. This therefore serves as a very useful tool especially in models where it is difficult to extract any cosmological solutions from directly solving the field equations such as in f(R) gravity. In the cases we have considered so far, one can for example write the equations \eqref{ftfriedmann} as an autonomous dynamical system using the variables \begin{equation}
\tilde{x} = -\frac{\ddot{f}}{H\dot{f}}\,, \quad\tilde{y} = \frac{f}{4H^2\dot{f}}\,, \quad\tilde{z} = \frac{3H^2+\dot{H}}{H^2}\,,
\end{equation}
These variables were considered in \cite{Ganiou:2018dta} and the authors considered the following scenarios: (i) absence of matter fluids and (ii) presence of dust and radiation components. Furthermore, the case when the parameter $m = -\frac{\ddot{H}}{H^3}$ takes on constant values $m = 0$ (quasi-de Sitter evolution) and $m = -\frac{9}{2}$ (matter dominated evolution) was explored. One can also write \eqref{ftbfriedmann} in the dynamical systems method using the phase space variables \begin{align}
    \tilde{x} &:= \frac{\dot{\phi}}{\sqrt{1+H^2}} \,, \quad \tilde{y} := \frac{V(\phi)}{6H^2} \,, \\
    \tilde{z} &:= \frac{(-T)^n}{1+H^2} \,, \quad \eta := \frac{H}{\sqrt{1+H^2}} \,.
\end{align}
The cosmological dynamics of $f(T,T_G)$ gravity was investigated in \cite{Kofinas:2014aka}. In particular, the model $f(T,T_G) = -T + \alpha_1 \sqrt{T^2 + \alpha_2 T_G}$, where $\alpha_{1,2} \neq 0$ are constants, was studied. In this case, the presence of a perfect dust fluid was assumed and the following dimensionless phase-space parameters were defined
\begin{equation}
\tilde{x} = \sqrt{1+\frac{2\alpha_2}{3}\left(1+\frac{\dot{H}}{H^2}\right)}\,, \quad \Omega_{\rm m} = \frac{\kappa^2 \rho_{\rm m}}{3H^2}\,.
\end{equation}
\\
\\
While we have briefly discussed several approaches to teleparallel gravity and their status quo in cosmology, what we are most interested in this work are the $f(T,\phi)$ models and we shall now discuss them in more detail.The action of TEGR (Teleparallel Equivalent of General Relativity) can be generalized to $f(T)$ gravity by introducing a scalar field $\phi$. The action, including matter and radiation, can be expressed as \cite{Gonzalez-Espinoza:2021mwr,Duchaniya:2022fmc}:

\begin{equation}\label{1}
    S = \int d^{4}x\, e[f(T,\phi) + P(\phi)X] + S_{m} + S_{r}\,,
\end{equation}

Here, $e = \det[e^A_{\mu}] = \sqrt{-g}$ represents the determinant of the tetrad field. The tetrad field, $e^{A}_{\mu}$, $A = 0,1,2,3$, is related to the metric tensor $g_{\mu \nu}$ and the Minkowski tangent space metric $\eta_{AB}$ as $g_{\mu \nu}=\eta_{AB} e_{\mu}^{A} e_{\nu}^{B}$, where $\eta_{AB}=(-1,1,1,1)$. The tetrad satisfies the orthogonality condition $e^{\mu}_Ae^B_{\mu}=\delta_A^B$, and the spin connection is denoted by ${\omega}^{A}_{{B\mu}}$. The function $f(T,\phi)$ represents an arbitrary function of the scalar field $\phi$ and the torsion scalar $T$, while $X= -\partial_{\mu} \phi \partial^{\mu} \phi/2$ represents the kinetic term of the field. This general action includes non-minimally coupled scalar-torsion gravity models with the coupling function $f(T, \phi)$, $f(T)$ gravity, and minimally coupled scalar field.
\\
\\
For a flat FLRW space-time background, the field equations derived from the action are \cite{Duchaniya:2022fmc}:

\begin{eqnarray}
    f(T,\phi) - P(\phi)X - 2Tf_{,T} &=& \rho_{m}+\rho_{r} \label{2} \\
    f(T,\phi) + P(\phi)X - 2Tf_{,T} - 4\dot{H}f_{,T} - 4H\dot{f}_{,T} &=& -p_{r} \label{3} \\
    -P_{,\phi}X - 3P(\phi)H\dot{\phi} - P(\phi)\ddot{\phi} + f_{,\phi} &=& 0 \label{4}
\end{eqnarray}

In these equations, the Hubble parameter is denoted as $H\equiv\frac{\dot{a}}{a}$, where an overdot represents a derivative with respect to cosmic time $t$. The energy density for matter and radiation are denoted as $\rho_{m}$ and $\rho_{r}$ respectively, and the pressure at the radiation era is $p_{r}$. The torsion scalar $T$ is given by $T=6H^{2}$. The non-minimal coupling function $f(T,\phi)$ is defined as \cite{Hohmann:2018rwf}:

\begin{equation}\label{5}
    f(T,\phi)=-\frac{T}{2\kappa^{2}}-G(T)-V(\phi)\,,
\end{equation}

where $V(\phi)$ is the scalar potential and $G(T)$ is an arbitrary function of the torsion scalar.

In the matter-dominated era, $\omega_{m}=\frac{p_{\rm m}}{\rho_{\rm m}} = 0$, and in the radiation era, $\omega_{r}=\frac{p_{r}}{\rho_{\rm r}} = 1/3$. In this case, Eqs.~\eqref{2}--\eqref{4} reduce to:

\begin{equation}
    \frac{3}{\kappa^{2}}H^{2} = P(\phi)X + V(\phi) - 2TG_{,T} + G(T) + \rho_{m} + \rho_{r}\,, \label{6}
\end{equation}
\begin{equation}
    -\frac{2}{\kappa^{2}}\dot{H} = 2P(\phi)X + 4\dot{H}(G_{T}+2TG_{,TT}) + \rho_{m} + \frac{4}{3}\rho_{r}\,, \label{7}  
\end{equation}
\begin{equation}
    P(\phi)\ddot{\phi} + P_{,\phi}(\phi)X + 3P(\phi)H\dot{\phi} + V_{,\phi}(\phi) = 0\,. \label{8}
\end{equation}

The modified Friedmann equations, taking into account dark energy, become:

\begin{eqnarray}
    \frac{3}{\kappa^{2}}H^{2} &=& \rho_{m} + \rho_{r} + \rho_{de}\,, \label{9} \\
    -\frac{2}{\kappa^{2}}\dot{H} &=& \rho_{m} + \frac{4}{3}\rho_{r} + \rho_{de} + p_{de}\,. \label{10}
\end{eqnarray}

Comparing Eq.~\eqref{6} with Eq.~\eqref{9}, and Eq.~\eqref{7} with Eq.~\eqref{10}, we can extract the energy density ($\rho_{de}$) and pressure ($p_{de}$) for the dark energy sector:

\begin{align}
    \rho_{de} &= P(\phi)X + V(\phi) - 2TG_{,T} + G(T)\,, \label{11} \\
    p_{de} &= P(\phi)X - V(\phi) + 2TG_{,T} - G(T) + 4\dot{H}(G_{,T}+2TG_{,TT})\,. \label{12}
\end{align}
For simplicity we set $P(\phi) = 1 $ and consider the well studied exponential potential, $V(\phi)=V_{0}e^{-\lambda\phi}$, where $\lambda$ is a constant. In order to proceed further and really carry out the analysis we want to, we need a form for $G(T)$ and here we will be considering two forms which were studied in \cite{Duchaniya:2022fmc} and will be carrying out the Goriely-Hyde analysis on both of these models \footnote{\textbf{It is worth discussing any effects of the separation of T from $\phi$ here in the action \eqref{5}. If one, for example, considers actions like those in \cite{Geng:2011ka} where terms like $T \phi^2 $ come into play then one could potentially expect that results on singularities could be affected in some ways, as one would not be wrong to think that such a coupling between the torsion scalar and field terms could have some significant outcomes. Although a definitive answer on this aspect would need one to do a proper analysis similar to the one we have done for the models considered in our paper, we do feel very interesting results may await an endeavour like this.} }.
\\
\\
\section{The Goriely-Hyde Procedure}

The Goriely-Hyde technique \cite{goriely2000necessary} offers an elegant method for identifying finite-time singularities in dynamical systems. The procedure can be summarized as follows:

\begin{itemize}
\item We begin by considering a dynamical system governed by $n$ differential equations given by:
\begin{equation}
\dot{x}{i} = f{i}(x),
\end{equation}
where $i = 1, 2, ..., n$. Here, $t$ represents time, but in quintessence models, it can be better represented as the number of e-foldings, denoted by $N$. We identify the parts of the equation $f_{i}$ that become significant as the system approaches the singularity. These significant parts are referred to as "dominant parts" \cite{goriely2000necessary}. Each dominant part represents a mathematically consistent truncation of the system, denoted as $\hat{f}{i}$. Consequently, the system can be expressed as:
\begin{equation}
\dot{x}{i} = \hat{f}_{i}(x).
\end{equation}
\item Without loss of generality, the variables $x_{i}$ near the singularity can be represented as:
\begin{equation}
x_{i} = a_{i} \tau^{p_{i}},
\end{equation}
where $\tau = t - t_{c}$, and $t_{c}$ is an integration constant. By substituting equation (4) into equation (3) and equating the exponents, we can determine the values of $p_{i}$ for different $i$, which collectively form the vector $\mathbf{p} = (p_{1}, p_{2}, ..., p_{n})$. Similarly, we calculate the values of $a_{i}$ to form the vector $\vec{a} = (a_{1}, a_{2}, ..., a_{n})$. It is worth noting that if $\vec{a}$ comprises solely real entries, it corresponds to finite-time singularities. On the other hand, if $\vec{a}$ contains at least one complex entry, it may lead to non-finite-time singularities. Each $(a_{i}, p_{i})$ set is referred to as a dominant balance of the system.

\item Next, we compute the Kovalevskaya matrix defined as:
\begin{equation}
R = \begin{pmatrix}
\frac{\partial f_{1}}{\partial x_{1}} & \frac{\partial f_{1}}{\partial x_{2}} & . & . & \frac{\partial f_{1}}{\partial x_{n}}\\
\frac{\partial f_{2}}{\partial x_{1}} & \frac{\partial f_{2}}{\partial x_{2}} & . & . & \frac{\partial f_{2}}{\partial x_{n}}\\
. & . & . & . & . \\
. & . & . & . & . \\
\frac{\partial f_{n}}{\partial x_{1}} & \frac{\partial f_{n}}{\partial x_{2}} & . & . & \frac{\partial f_{n}}{\partial x_{n}}\\
\end{pmatrix} -  \begin{pmatrix}
p_{1} & 0 & . & . & 0 \\
0 & p_{2} & . & . & 0 \\
. & . & . & . & . \\
. & . & . & . & . \\
0 & 0 & . & . & p_{n} \\
\end{pmatrix}.
\end{equation}

After obtaining the Kovalevskaya matrix, we evaluate it for different dominant balances and determine the eigenvalues. If the eigenvalues take the form $(-1, r_{2}, r_{3}, ..., r_{n})$, where $r_{2}, r_{3}, ... > 0$, then the singularity is regarded as general and will occur regardless of the initial conditions of the system. Conversely, if any of the eigenvalues $r_{2}, r_{3}, ...$ are negative, the singularity is considered local and will only occur for certain sets of initial conditions.
\end{itemize}
\section{Goriely-Hyde analysis of $f(T,\phi) $ models }

\subsection{Model I}\label{sec:model_1}

In the first model, we consider a specific form for $G(T)$ as given in  \cite{2011JCAP...07..015Z,Duchaniya:2022fmc}:
\begin{equation}\label{13}
    G(T) = \beta T \ln \left(\frac{T}{T_{0}}\right)\,,
\end{equation}
where $\beta$ is a constant and $T_0$ represents the value of $T$ at the initial epoch. This model, which has been investigated in \cite{2011JCAP...07..015Z}, exhibits physically favorable critical points and offers an interesting approach for modeling the evolution of the Universe. By substituting this expression into Eqs.~\eqref{11}--\eqref{12}, the effective dark energy density and pressure terms are reduced to:
\begin{equation}
    \rho_{de} = \frac{\dot{\phi}^{2}}{2} + V(\phi) - 6\beta H^{2} \ln \left(\frac{6 H^{2}}{T_{0}}\right) - 12 H^{2} \beta\,, \label{14}  
\end{equation}
\begin{multline}
    p_{de} = \frac{\dot{\phi}^{2}}{2} - V(\phi) + 6\beta H^{2} \ln \left(\frac{6H^{2}}{T_{0}}\right) + 12 H^{2} \beta + \\ 4 \dot{H} \left(\beta \ln \left(\frac{6H^{2}}{T_{0}} \right)+3 \beta \right)\,. \label{15} 
\end{multline}
In order to analyze the dynamics of the scalar-torsion $f(T,\phi)$ gravity model, \cite{Duchaniya:2022fmc} introduced a set of dimensionless phase space variables to represent the system in an autonomous form. These variables are defined as \footnote{Note that it is not necessary that the variables will be defined in this same way for all cosmological paradigms, as we shall see later in the paper too. In fact, one can use different variables for the same paradigm too if required or wished for. See, for example, \cite{halliwell1987scalar,copeland1998exponential,faraoni2013scalar} for extended discussions on the same} follows:
\begin{multline}
    x = \frac{\kappa\dot{\phi}}{\sqrt{6}H}\,, \hspace{1cm} y = \frac{\kappa\sqrt{V}}{\sqrt{3}H}\,, \\ \hspace{1cm}
    z = -4 \beta \kappa^{2}\,, \hspace{1cm}
    u = -2 \beta \ln \left(\frac{T}{T_{0}}\right)\kappa^{2} \\
    \rho = \frac{\kappa\sqrt{\rho_{r}}}{\sqrt{3}H}\,, \hspace{1cm} 
    \lambda = -\frac{V_{,\phi}(\phi)}{\kappa V(\phi)}\,, \hspace{1cm} 
    \Theta = \frac{V(\phi)\,, V_{,\phi \phi}}{V_{,\phi}(\phi)^{2}}\,. \label{17}
\end{multline}
These dimensionless variables allow for a simplified representation of the system's dynamics and facilitate the analysis of the scalar-torsion $f(T,\phi)$ gravity model. Using these variables, one can finally write the cosmological equations of this model as a dynamical system as follows : \begin{align*}
    \frac{dx}{dN}&=-\frac{x\rho ^2-3 x\left(u-x^2+y^2+z-1\right)}{2 u+3 z-2}-3 x+\sqrt{\frac{3}{2}} \lambda  y^2\,,  \\ 
    \frac{dy}{dN}&=\frac{-y \rho ^2+3y \left(u-x^2+y^2+z-1\right)}{2 u+3 z-2}-\sqrt{\frac{3}{2}} \lambda y x\,, \ \\
    \frac{du}{dN}&=\frac{z \rho ^2-3 z \left(u-x^2+y^2+z-1\right)}{2 u+3 z-2}\,,  \\ 
    \frac{d\rho}{dN}&=-\frac{\rho \left(\rho ^2+u+3 x^2-3 y^2+3 z-1\right)}{2 u+3 z-2}\,, \\
    \frac{dz}{dN}&=0\,,  \\
    \frac{d\lambda}{dN}&= -\sqrt{6}(\Theta-1)x \lambda^{2}\,. 
\end{align*}

For our analysis , we would be considering $\lambda $ to be a constant \textbf{which is not equal to zero} (which would again mean that we are considering an exponential potential form as we remarked earlier). Furthermore, we consider that $ 2u >> 3z -2 $ ( which can be justified considering the forms of z and u we have described earlier ). This would allow us to write the dynamical equations as \begin{equation}
    \frac{dx}{dN} = -\frac{x\rho^2-3x(u-x^2+y^2+z-1)}{2u} - 3x + \sqrt{\frac{3}{2}}\lambda y^2  \label{18}
\end{equation}
\begin{equation}
    \frac{dy}{dN} = \frac{-y\rho^2+3y(u-x^2+y^2+z-1)}{2u} - \sqrt{\frac{3}{2}}\lambda yx \label{19}
\end{equation}
\begin{equation}
    \frac{du}{dN} = \frac{z\rho^2-3z(u-x^2+y^2+z-1)}{2u} \label{20}
\end{equation}
\begin{equation}
    \frac{d\rho}{dN} = -\frac{\rho(\rho^2+u+3x^2-3y^2+3z-1)}{2u} \label{21}
\end{equation}
\begin{equation}
    \frac{dz}{dN} = 0 \label{22}
\end{equation}
\begin{equation}
    \frac{d\lambda}{dN} = 0 \label{23}
\end{equation}
Now we are in the right position to start off our singularity analysis. The first truncation that we consider is given by \begin{equation}
         \hat{f} = \begin{pmatrix}
         \sqrt{\frac{3}{2}} \lambda  y^2  \\
          3 y^3 / 2u  \\ 
          3z/ 2u \\
          \rho^3 / 2u
         \end{pmatrix}
         \end{equation} 
Using the ansatz of the Goriely-Hyde method, we get the exponents to be $p = (1/2,-1/4,1/2,-1/4) $ from which we can get the dominant balances to be
\begin{equation}. \label{a1}
    \begin{aligned}
      a_{1} = \left( -\frac{\lambda  z}{\sqrt{2}} , \frac{i \sqrt[4]{z}}{\sqrt{2} \sqrt[4]{3}} , \sqrt{3 z} , \frac{\sqrt[4]{3} \sqrt[4]{z}}{\sqrt{2}} \right)   \\[10pt]
      a_{2} = \left( -\frac{\lambda  z}{\sqrt{2}} , - \frac{i \sqrt[4]{z}}{\sqrt{2} \sqrt[4]{3}} , \sqrt{3 z} , \frac{\sqrt[4]{3} \sqrt[4]{z}}{\sqrt{2}} \right)   \\[10pt]
      a_{3} = \left( -\frac{\lambda  z}{\sqrt{2}} , \frac{i \sqrt[4]{z}}{\sqrt{2} \sqrt[4]{3}} , - \sqrt{3 z} , \frac{\sqrt[4]{3} \sqrt[4]{z}}{\sqrt{2}} \right)   \\[10pt]
      a_{4} = \left( -\frac{\lambda  z}{\sqrt{2}} , \frac{i \sqrt[4]{z}}{\sqrt{2} \sqrt[4]{3}} , \sqrt{3 z} ,- \frac{\sqrt[4]{3} \sqrt[4]{z}}{\sqrt{2}} \right)   \\[10pt]
      a_{5} = \left( -\frac{\lambda  z}{\sqrt{2}} , \frac{i \sqrt[4]{z}}{\sqrt{2} \sqrt[4]{3}} , - \sqrt{3 z} , - \frac{\sqrt[4]{3} \sqrt[4]{z}}{\sqrt{2}} \right)   \\[10pt]
      a_{6} = \left( -\frac{\lambda  z}{\sqrt{2}} , -\frac{i \sqrt[4]{z}}{\sqrt{2} \sqrt[4]{3}} , -\sqrt{3 z} , \frac{\sqrt[4]{3} \sqrt[4]{z}}{\sqrt{2}} \right)   \\[10pt]
      a_{7} = \left( -\frac{\lambda  z}{\sqrt{2}} , -\frac{i \sqrt[4]{z}}{\sqrt{2} \sqrt[4]{3}} , \sqrt{3 z} , -\frac{\sqrt[4]{3} \sqrt[4]{z}}{\sqrt{2}} \right)   \\[10pt]
      a_{8} = \left( -\frac{\lambda  z}{\sqrt{2}} ,- \frac{i \sqrt[4]{z}}{\sqrt{2} \sqrt[4]{3}} , - \sqrt{3 z} , - \frac{\sqrt[4]{3} \sqrt[4]{z}}{\sqrt{2}} \right) 
    \end{aligned}
\end{equation}

We can now write the Kovalevskaya matrix to be \begin{equation} . \label{a2}
    R = \left(
\begin{array}{cccc}
 -\frac{1}{2} & \sqrt{6} \lambda  y & 0 & 0 \\
 0 & \frac{9 y^2}{2 u}+\frac{1}{4} & -\frac{3 y^3}{2 u^2} & 0 \\
 0 & 0 & -\frac{3 z}{2 u^2}-\frac{1}{2} & 0 \\
 0 & 0 & \frac{\rho ^3}{2 u^2} & \frac{1}{4}-\frac{3 \rho ^2}{2 u} \\
\end{array}
\right)
\end{equation}

Using the dominant balances we introduced in \eqref{a1}, we can now plug them into the Kovalevskaya matrix \eqref{a2} to get the eigenvalues to be \begin{equation}
    r = (-1,-1/2,-1/2,-1/2) 
\end{equation} 
We note that all the other eigenvalues besides the initial -1 are also negative, which means that according to the Goriely-Hyde method the singularities of this system with regards to this truncation can occur only for a limited set of initial conditions. Coupled with the fact that the dominant balances \eqref{a1} have complex entries, this would mean that this truncation tells us that the singularities for this system may occur in non-finite time.
\\
\\
The second truncation that we consider is given by \begin{equation}
         \hat{f} = \begin{pmatrix}
         3 x^3 / 2u  \\
         -\sqrt{\frac{3}{2}} \lambda  x y  \\ 
           \rho ^2 z /2 u  \\
           3 \rho  y^2 / 2 u  
         \end{pmatrix}
         \end{equation} 
Using the ansatz of the Goriely-Hyde method, we get the exponents to be $p = (-1,-1,-1,-3/2) $ from which we can get the dominant balances to be 
\begin{equation}. \label{a3}
    \begin{aligned}
      a_{1} = \left( \frac{\sqrt{\frac{2}{3}}}{\lambda } , \frac{1}{\lambda } , -\frac{1}{\lambda ^2} , \frac{i \sqrt{\frac{2}{z}}}{\lambda ^2} \right)   \\[10pt]
      a_{2} = \left( - \frac{\sqrt{\frac{2}{3}}}{\lambda } , \frac{1}{\lambda } , -\frac{1}{\lambda ^2} , \frac{i \sqrt{\frac{2}{z}}}{\lambda ^2} \right)   \\[10pt]
      a_{3} = \left( \frac{\sqrt{\frac{2}{3}}}{\lambda } , - \frac{1}{\lambda } , -\frac{1}{\lambda ^2} , \frac{i \sqrt{\frac{2}{z}}}{\lambda ^2} \right)   \\[10pt]
      a_{4} = \left( \frac{\sqrt{\frac{2}{3}}}{\lambda } , \frac{1}{\lambda } , -\frac{1}{\lambda ^2} , - \frac{i \sqrt{\frac{2}{z}}}{\lambda ^2} \right)   \\[10pt]
      a_{5} = \left( - \frac{\sqrt{\frac{2}{3}}}{\lambda } ,- \frac{1}{\lambda } , -\frac{1}{\lambda ^2} , \frac{i \sqrt{\frac{2}{z}}}{\lambda ^2} \right)  \\[10pt]
      a_{6} = \left(- \frac{\sqrt{\frac{2}{3}}}{\lambda } , \frac{1}{\lambda } , -\frac{1}{\lambda ^2} , - \frac{i \sqrt{\frac{2}{z}}}{\lambda ^2} \right)   \\[10pt]
      a_{7} = \left( \frac{\sqrt{\frac{2}{3}}}{\lambda } , - \frac{1}{\lambda } , -\frac{1}{\lambda ^2} , -\frac{i \sqrt{\frac{2}{z}}}{\lambda ^2} \right)   \\[10pt]
      a_{8} = \left(- \frac{\sqrt{\frac{2}{3}}}{\lambda } , - \frac{1}{\lambda } , -\frac{1}{\lambda ^2} , - \frac{i \sqrt{\frac{2}{z}}}{\lambda ^2} \right) 
    \end{aligned}
\end{equation}
We again see that the balances have complex entries \footnote{At this point we would like to highlight that complex entries in $\mathbf{\hat{a}}$  observed for the previous truncation and this one ( and which will be observed for Model II as well for a few truncations) are completely consistent with the fact that the system  consists of expansion normalized variables which are real. As mentioned in section 2, complex entries for various $\mathbf{a}$ suggest that the singularities will be non-finite time in nature and hence these quantities taking up complex values is consistent with the analysis as shown in \cite{goriely2000necessary}. Similar case has been for various cosmological systems (for example, see \cite{odintsov2018dynamical,odintsov2019finite})} while the the Kovalevskaya matrix can be written as \begin{equation} . \label{a4}
    R = \left(
\begin{array}{cccc}
 \frac{9 x^2}{2 u}+1 & 0 & -\frac{3 x^3}{2 u^2} & 0 \\
 -\sqrt{\frac{3}{2}} \lambda  y & 1-\sqrt{\frac{3}{2}} \lambda  x & 0 & 0 \\
 0 & 0 & 1-\frac{\rho ^2 z}{2 u^2} & \frac{\rho  z}{u} \\
 0 & \frac{3 \rho  y}{u} & -\frac{3 \rho  y^2}{2 u^2} & \frac{3 y^2}{2 u}+\frac{3}{2} \\
\end{array}
\right)
\end{equation}

Using the dominant balances we introduced in \eqref{a3}, we can now plug them into the Kovalevskaya matrix \eqref{a4} to get the eigenvalues to be \begin{equation}
    r \sim (-1,1.27,-1.5,1.27) 
\end{equation} 
We note that as one of the other eigenvalues (-1.5) besides the initial -1 is also negative, according to the Goriely-Hyde method the singularities of this system with regards to this truncation can occur only for a limited set of initial conditions. Coupled with the fact that the dominant balances \eqref{a3} have complex entries, this would mean that this truncation tells us that the singularities for this system may occur in non-finite time.
\\
\\
The third truncation that we consider is given by \begin{equation}
         \hat{f} = \begin{pmatrix}
         - \rho^2 x / 2u  \\
         - \rho^2 y / 2u  \\ 
           3 x^2 z / 2u  \\
           -3 \rho z / 2u 
         \end{pmatrix}
         \end{equation} 
Using the ansatz of the Goriely-Hyde method, we get the exponents to be $p = (1/2,1/2,1,1/4) $ from which we can get the dominant balances to be
\begin{equation}. \label{a5}
    \begin{aligned}
      a_{1} = \left( 2 \sqrt{6z} , 2 \sqrt{6z} , -6z , 2 \sqrt{3z} \right)   \\[10pt]
      a_{2} = \left( - 2 \sqrt{6z} , 2 \sqrt{6z} , -6z , 2 \sqrt{3z} \right)   \\[10pt]
      a_{3} = \left(  2 \sqrt{6z} , -2 \sqrt{6z} , -6z , 2 \sqrt{3z} \right)   \\[10pt]
      a_{4} = \left( 2 \sqrt{6z} , 2 \sqrt{6z} , -6z , -2 \sqrt{3z} \right)   \\[10pt]
      a_{5} = \left( -2 \sqrt{6z} ,- 2 \sqrt{6z} , -6z , 2 \sqrt{3z} \right)  \\[10pt]
      a_{6} = \left( -2 \sqrt{6z} , 2 \sqrt{6z} , -6z , -2 \sqrt{3z} \right)   \\[10pt]
      a_{7} = \left( 2 \sqrt{6z} , -2 \sqrt{6z} , -6z , -2 \sqrt{3z} \right)   \\[10pt]
      a_{8} = \left(- 2 \sqrt{6z} , -2 \sqrt{6z} , -6z , -2 \sqrt{3z} \right) 
    \end{aligned}
\end{equation}

We can now write the Kovalevskaya matrix to be \begin{equation} . \label{a6}
    R = \left(
\begin{array}{cccc}
 -\frac{\rho ^2}{2 u}-\frac{1}{2} & 0 & \frac{\rho ^2 x}{2 u^2} & -\frac{\rho  x}{u} \\
 0 & -\frac{\rho ^2}{2 u}-\frac{1}{2} & \frac{\rho ^2 y}{2 u^2} & -\frac{\rho  y}{u} \\
 \frac{3 x z}{u} & 0 & -\frac{3 x^2 z}{2 u^2}-1 & 0 \\
 0 & 0 & \frac{3 \rho  z}{2 u^2} & -\frac{3 z}{2 u}-\frac{1}{4} \\
\end{array}
\right)
\end{equation}

Using the dominant balances we introduced in \eqref{a5}, we can now plug them into the Kovalevskaya matrix \eqref{a6} to get the eigenvalues to be \begin{equation}
    r \sim (-1,1/2,-0.1,-0.1) 
\end{equation} 
We note that as two of the other eigenvalues besides the initial -1 are also negative, according to the Goriely-Hyde method the singularities of this system with regards to this truncation can occur only for a limited set of initial conditions. But in this case we see something which we didn't in the previous two truncations ; the dominant balances \eqref{a5} do not have complex entries. This means that this truncation tells us that it is definitely possible to have singularities occuring in finite-time for this particular model. While we can go on and evaluate more truncations, we find that in no other truncation would we see something which we have not observed already in these three truncations. Namely that there is no truncation for this system for which the eigenvalues besides -1 are all positive and so it does seem like for this model the singularities will not be general and can only happen for a limited set of initial conditions.
\subsection{Model II}\label{sec:model_2}
In this scenario, we consider the function $G(T)$ to be of the form $G(T) = T + \alpha T^{2}$, where $\alpha$ is a constant \cite{fortes2022solving}. This represents a slight extension beyond the Teleparallel Equivalent of General Relativity (TEGR), as $\alpha=0$ corresponds to the TEGR model. For this particular $G(T)$, Eqs. \eqref{17}-\eqref{18} can be expressed as follows:

\begin{eqnarray}
    \rho_{de} &=& \frac{\dot{\phi^{2}}}{2}+V(\phi)-T(1+3T \alpha)\,, \label{24}\\
    p_{de} &=& \frac{\dot{\phi^{2}}}{2}-V(\phi)+T(1+3 T \alpha)+4 \dot{H}(1+6 T \alpha)\,. \label{25}
\end{eqnarray}

To establish an independent dynamical system, we introduce dimensionless variables defined as:

\begin{multline}
    x = \frac{\kappa\dot{\phi}}{\sqrt{6}H} \quad y = \frac{\kappa\sqrt{V}}{\sqrt{3}H} \\ \quad
    z = -2 \kappa^{2} \quad
    u = -36 H^{2} \alpha \kappa^{2} \\
    \rho = \frac{\kappa\sqrt{\rho_{r}}}{\sqrt{3}H} \quad 
    \lambda = -\frac{V_{,\phi}(\phi)}{\kappa V(\phi)} \quad 
    \Theta = \frac{V(\phi) V_{,\phi \phi}}{V_{,\phi}(\phi)^{2}} \label{26}
\end{multline}

Consequently, the corresponding dynamical system can be obtained as,
\begin{align*}
    \frac{dx}{dN}&=-\frac{x \left(\rho ^2-3 \left(u-x^2+y^2+z-1\right)\right)}{2 (2 u+z-1)}-3 x+\sqrt{\frac{3}{2}} \lambda y^2\,,  \\ 
    \frac{dy}{dN}&=-\frac{1}{2} y \left(\frac{\rho ^2-3 \left(u-x^2+y^2+z-1\right)}{2 u+z-1}+\sqrt{6} \lambda x\right)\,,  \\ 
    \frac{du}{dN}&=\frac{u \left(\rho ^2-3 \left(u-x^2+y^2+z-1\right)\right)}{2 (2 u+z-1)}\,,  \\
    \frac{d\rho}{dN}&=-\frac{\rho \left(\rho ^2+5 u+3 x^2-3 y^2+z-1\right)}{2 (2 u+z-1)}\,,  \\ 
    \frac{dz}{dN}&=0\,,  \\
    \frac{d\lambda}{dN}&= -\sqrt{6}(\Theta-1)x \lambda^{2}\,.
\end{align*}

We again consider $\lambda$ to be a constant here, which would mean that we are interested in exponential potentials. Furthermore, we assume that $2u >> z -1 $ which is again not hard to justify considering the definitions of these quantities in  \eqref{26}. By taking these considerations into account, the dynamical system takes the form \begin{align}
    \frac{dx}{dN}&=-\frac{x \left(\rho ^2-3 \left(u-x^2+y^2+z-1\right)\right)}{4u}-3 x+\sqrt{\frac{3}{2}} \lambda y^2\,, \label{28} \\ 
    \frac{dy}{dN}&=-\frac{1}{2} y \left(\frac{\rho ^2-3 \left(u-x^2+y^2+z-1\right)}{2u}+\sqrt{6} \lambda x\right)\,, \label{29} \\ 
    \frac{du}{dN}&=\frac{u \left(\rho ^2-3 \left(u-x^2+y^2+z-1\right)\right)}{4u}\,, \label{30} \\
    \frac{d\rho}{dN}&=-\frac{\rho \left(\rho ^2+5 u+3 x^2-3 y^2+z-1\right)}{4u}\,, \label{31} \\ 
    \frac{dz}{dN}&=0\,, \label{32} \\
    \frac{d\lambda}{dN}&= 0\,. \label{33}
\end{align}
We can now start with the Goriely-Hyde analysis of this system, with the first truncation that we consider being \begin{equation}
         \hat{f} = \begin{pmatrix}
         \sqrt{\frac{3}{2}} \lambda  y^2  \\
          -y \rho^2 / 4u  \\ 
          3 y^2 \\
          -3 \rho x^2 / 4u
         \end{pmatrix}
         \end{equation} 
Using the ansatz of the Goriely-Hyde method, we get the exponents to be $p=(-1,-1,-1,-1)$ from which we can get the dominant balances to be
\begin{equation}. \label{a9}
    \begin{aligned}
      a_{1} = \left( \frac{1}{\lambda} \sqrt{\frac{2}{3}} , \frac{1}{\lambda} \sqrt{\frac{2}{3}} , \frac{1}{2 \lambda^2 } , \frac{\sqrt{2}}{\lambda} \right)   \\[10pt]
      a_{2} = \left( - \frac{1}{\lambda} \sqrt{\frac{2}{3}} , \frac{1}{\lambda} \sqrt{\frac{2}{3}} , \frac{1}{2 \lambda^2 } , \frac{\sqrt{2}}{\lambda} \right)    \\[10pt]
      a_{3} = \left( \frac{1}{\lambda} \sqrt{\frac{2}{3}} , - \frac{1}{\lambda} \sqrt{\frac{2}{3}} , \frac{1}{2 \lambda^2 } , \frac{\sqrt{2}}{\lambda} \right)   \\[10pt]
      a_{4} = \left( \frac{1}{\lambda} \sqrt{\frac{2}{3}} , \frac{1}{\lambda} \sqrt{\frac{2}{3}} , \frac{1}{2 \lambda^2 } , - \frac{\sqrt{2}}{\lambda} \right)    \\[10pt]
      a_{5} = \left(- \frac{1}{\lambda} \sqrt{\frac{2}{3}} , - \frac{1}{\lambda} \sqrt{\frac{2}{3}} , \frac{1}{2 \lambda^2 } , \frac{\sqrt{2}}{\lambda} \right)   \\[10pt]
      a_{6} = \left(- \frac{1}{\lambda} \sqrt{\frac{2}{3}} , \frac{1}{\lambda} \sqrt{\frac{2}{3}} , \frac{1}{2 \lambda^2 } , -\frac{\sqrt{2}}{\lambda} \right)    \\[10pt]
      a_{7} = \left( \frac{1}{\lambda} \sqrt{\frac{2}{3}} , -\frac{1}{\lambda} \sqrt{\frac{2}{3}} , \frac{1}{2 \lambda^2 } , -\frac{\sqrt{2}}{\lambda} \right)    \\[10pt]
      a_{8} = \left(- \frac{1}{\lambda} \sqrt{\frac{2}{3}} , - \frac{1}{\lambda} \sqrt{\frac{2}{3}} , \frac{1}{2 \lambda^2 } , - \frac{\sqrt{2}}{\lambda} \right)  
    \end{aligned}
\end{equation}

We can now write the Kovalevskaya matrix to be \begin{equation} . \label{a10}
    R = \left(
\begin{array}{cccc}
 1 & \sqrt{6} \lambda  y & 0 & 0 \\
 0 & 1-\frac{\rho ^2}{4 u} & \frac{\rho ^2 y}{4 u^2} & -\frac{\rho  y}{2 u} \\
 0 & 6 y & 1 & 0 \\
 -\frac{3 \rho  x}{2 u} & 0 & \frac{3 \rho  x^2}{4 u^2} & 1-\frac{3 x^2}{4 u} \\
\end{array}
\right)
\end{equation}

Using the dominant balances we introduced in \eqref{a9}, we can now plug them into the Kovalevskaya matrix \eqref{a10} to get the eigenvalues to be \begin{equation}
    r \sim (-1,1,-2 \sqrt{2},2 \sqrt{2}) 
\end{equation} 

As we have one of the eigenvalues besides -1 also being negative, this truncation tells us that the singularities that could appear for this model would also only be occuring for a limited set of initial conditions for the variables. Furthermore given that the dominant balances \eqref{a9} all have real entries then this would mean that the singularities only appear in finite time. 
\\
\\
The second truncation that we would be considering is given by \begin{equation}
         \hat{f} = \begin{pmatrix}
         \sqrt{\frac{3}{2}} \lambda  y^2  \\
          -y \rho^2 / 4u  \\ 
          3 y^2 \\
          -3 \rho x^2 / 4u
         \end{pmatrix}
         \end{equation} 
Using the ansatz of the Goriely-Hyde method, we get the exponents to be $p = (-1,-1,-1,-1) $ from which we can get the dominant balances to be
\begin{equation}. \label{a13}
    \begin{aligned}
      a_{1} = \left( \frac{1}{\lambda} \sqrt{\frac{2}{3}} , \frac{\sqrt{2} i}{\lambda} , - \frac{1}{2 \lambda^2 } , \frac{\sqrt{2} i}{\lambda} \right)   \\[10pt]
      a_{2} = \left( \frac{1}{\lambda} \sqrt{\frac{2}{3}} ,- \frac{\sqrt{2} i}{\lambda} , - \frac{1}{2 \lambda^2 } , \frac{\sqrt{2} i}{\lambda} \right)   \\[10pt]
      a_{3} = \left( \frac{1}{\lambda} \sqrt{\frac{2}{3}} , \frac{\sqrt{2} i}{\lambda} , - \frac{1}{2 \lambda^2 } , -\frac{\sqrt{2} i}{\lambda} \right)  \\[10pt]
      a_{4} = \left(\frac{1}{\lambda} \sqrt{\frac{2}{3}} , -\frac{\sqrt{2} i}{\lambda} , - \frac{1}{2 \lambda^2 } , -\frac{\sqrt{2} i}{\lambda} \right)  
    \end{aligned}
\end{equation}

We can now write the Kovalevskaya matrix to be \begin{equation} . \label{a14}
    R = \left(
\begin{array}{cccc}
 1-\frac{\rho ^2}{4 u} & 0 & \frac{\rho ^2 x}{4 u^2} & -\frac{\rho  x}{2 u} \\
 -\sqrt{\frac{3}{2}} \lambda  y & 1-\sqrt{\frac{3}{2}} \lambda  x & 0 & 0 \\
 \frac{3 x}{2} & 0 & 1 & 0 \\
 0 & \frac{3 \rho  y}{2 u} & -\frac{3 \rho  y^2}{4 u^2} & \frac{3 y^2}{4 u}+1 \\
\end{array}
\right)
\end{equation}

Using the dominant balances we introduced in \eqref{a13}, we can now plug them into the Kovalevskaya matrix \eqref{a14} to get the eigenvalues to be \begin{equation}
    r \sim (-1,1.6,3.7,-0.2) 
\end{equation} 
We again see that there are eigenvalues besides -1 which are negative, which again suggests that the model may not have general singularities. Furthermore this truncation also suggests that singularities could take place in non-finite time as shown by the complex entries in the dominant balance \eqref{a13}. While we can again go on for more truncations, what we have found out is that the other truncations do not offer anything new other than what we have seen so far. Namely, no truncation suggests that the model can allow for general singularities and so we are not going to be evaluating for more truncations here.
\section{Physical classification of the singularities}

Until now, we have discussed the singularity structure within the dark energy scenario from a dynamical perspective. However, it is insufficient to merely acknowledge the existence of singularities in this system from a physical standpoint. Thus, it becomes necessary to appropriately classify the potential types of singularities that could occur in this model. Various types of physical singularities for cosmology at a specific time $t = t_{s}$, where $t_{s}$ represents the occurrence of the singularities, can be classified as follows \cite{Nojiri:2005sx,Fernandez-Jambrina:2010ngm}:

\begin{itemize}
\item Type I ("Big Rip"): In this case, the scale factor $a$, effective energy density $\rho_{\text{eff}}$, and effective pressure density $p_{\text{eff}}$ diverge.
\item Type II ("Sudden/Quiescent singularity"): In this case, $p_{\text{eff}}$ diverges, as well as the derivatives of the scale factor beyond the second derivative.
\item Type III ("Big Freeze"): In this case, the derivative of the scale factor from the first derivative onwards diverges.
\item Type IV ("Generalized sudden singularities"): In this case, the derivative of the scale factor diverges from a derivative higher than the second.
\end{itemize}

Among these classifications, Type I singularities are considered strong singularities since they have the ability to distort finite objects, while singularities of Type II, Type III, and Type IV are regarded as weak singularities as they cannot be perceived as either the beginning or the end of the universe. Although there are other minor types of singularities, such as Type V singularities or "w" singularities, we will focus solely on Type I to Type IV singularities here. The most general form of the Hubble parameter for investigating singularities within the aforementioned classified types is expressed as \cite{odintsov2019finite}:

\begin{equation} \label{a11}
H(t) = f_{1}(t) + f_{2}(t)(t - t_{s})^{\mathbf{\epsilon}}
\end{equation}

Here, $f_{1}(t)$ and $f_{2}(t)$ are assumed to be nonzero regular functions at the time of the singularity, and similar conditions apply to their derivatives up to the second order. Additionally, $\mathbf{\epsilon}$ is a real number. It is not mandatory for the Hubble parameter (34) to be a solution to the field equations; however, we will consider this case and explore the implications of this assumption on the singularity structure based on our dynamic analysis. First, we observe that none of the variables $x$, $y$, or $z$ as defined in (10) can ever become singular for any cosmic time value. The singularities that can occur considering the Hubble parameter as defined in (34) are as follows:

\begin{itemize}
\item For $\epsilon < -1$, a big rip singularity occurs.
\item For $-1 < \epsilon < 0$, a Type III singularity occurs.
\item For $0 < \\epsilon < 1$, a Type II singularity occurs.
\item For $\epsilon > 1$, a Type IV singularity occurs.
\end{itemize}

Another ansatz useful for classifying singularities was introduced in \cite{odintsov2022did} whereby the scale factor was written as:
\begin{equation} \label{a12}
a(t) = g(t) (t-t_{s})^{\epsilon} + f(t)
\end{equation}
where $g(t)$ and $f(t)$ and all their higher-order derivatives with respect to cosmic time are smooth functions of the cosmic time. For this ansatz, according to the values of the exponent $\epsilon$, one can have the following singularities:

\begin{itemize}
\item For $\epsilon < 0$, a Type I singularity occurs.
\item For $0 < \epsilon < 1$, a Type III singularity develops.
\item For $1 < \epsilon < 2$, a Type II singularity occurs.
\item For $\epsilon > 2$, a Type IV singularity occurs.
\end{itemize}

Again, it is not mandatory for the scale factor in equation \eqref{a12} to necessarily be a solution to the field equations, but we would like to consider this and equation \eqref{a11} in order to gain a well-motivated understanding of the types of cosmological singularities we can encounter in the various models we have discussed so far.
\\
\\
To proceed further, we need to express the expansion normalized variables that we defined for both models in terms of the Hubble parameter alone. To do this, we realize that we need to express the potential and the derivative of the field parameter in each case in terms of the Hubble parameter as these are the quantities on which the expansion normalized variables really depend in both the scenario ( in this scenario we are talking about representing the x and y variables in both cases in terms of the Hubble parameter). For the model $G(T) = \beta T \ln \left(\frac{T}{T_{0}}\right)$ \eqref{13}, we have \begin{equation}
   \dot{\phi}^2_{\beta} =  \frac{2 \dot{H}}{\kappa^2} - \rho_{m} - \frac{4}{3} \rho_{r} - 4 \dot{H} \Bigg[ 
 \beta \ln \left( \frac{6 H^2}{T_{0}} \right) + 3 \beta \Bigg] 
\end{equation}  
While the potential for this case be written as \begin{equation}
   V_{\beta} =  \frac{\dot{H} \left(6 \beta +2 \beta  \ln \left(\frac{6 H^2}{T_{0}}\right)+\frac{1}{\kappa^2}\right)-\frac{\rho_{m}}{2}+\rho_{r}}{H^2}+3 \left(4 \beta +2 \beta  \ln \left(\frac{6 H^2}{T_{0}}\right)+\frac{1}{\kappa^2}\right)
\end{equation}
For the model $G(T) = T + \alpha T^{2}$, we have the same quantities to be \begin{equation}
    \dot{\phi}^2_{\alpha} = \dot{H} \left(-24 \alpha  H^2-\frac{2}{\mathbf{\kappa}^2}-4\right)-\rho_{m}-4 \rho_{r}
\end{equation}
\begin{equation}
    V_{\alpha} = H^2 \left(12 \alpha  \dot{H} +\frac{3}{\kappa^2}+1\right)+\left(\frac{1}{\kappa^2}+2\right) \dot{H} +3 \alpha  H^4-\frac{\rho_{m}}{2}+\rho_{r}
\end{equation}
Using these one can express the dynamical variables used in the Goriely-Hyde analysis of both the models (\eqref{17},\eqref{26}) completely in terms of the Hubble parameter ( we will not write the variables out explicitly here as they have quite long expressions) and now we can use both the ansatz \eqref{a11}-\eqref{a12} to see under what conditions will the variables blow up. Remember that we do not want the dynamical variables to blow up and the values of the exponents of the ansatz for which they do not blow up will tell us the singularities which will be possible for these models. The interesting conclusion that actually comes out when one puts both the ansatz into the dynamical variables is that only Type IV singularities are possible for both models. None of Type I, Type II or Type III singularities can occur for any of the models for any of the ansatz \eqref{a11}-\eqref{a12} while Type IV singularities do take place for both the models, for any of the ansatz'. This is quite an interesting behaviour which to the best of our knowledge has only been shown in $f(T,\phi) $ theories, in that one is only observing Type IV singularities for both of the models considered. This leads one to speculate the possibility that $f(T,\phi) $ gravity may be better suited for cosmology than some of their other modified gravity counterparts as the theory is only admitting weak singularities. Furthermore, given the analysis from the Goriely-Hyde procedure, one is lead to conclude that such singularities can only occur for a limited set of initial conditions and may occur in finite or even non-finite time. 
\section{Concluding remarks}
In this paper, we have considered a well studied formulation of teleparallel dark energy in the form of $f(T,\phi) $ gravity, where the scalar field drives the expansion of the universe. We considered two particular well studied models of this theory and probed cosmological singularities for both the scenarios. For this endeavor, we used a method pioneered by the works of Odintsov in recent years, in which we applied the Goriely-Hyde procedure to the various dynamical systems by which the cosmological equations of these three models could be described. This allowed us to make predictions about whether singularities in these scenarios would be strongly dependent on initial physical conditions and whether they could happen in finite or nonfinite times. After this, we employed two very well-motivated ansatz' for the Hubble parameter and the scale factor to reach 
the conclusion that one can only have Type IV singularities for both of the models considered in our work, that too only for a limited set of initial conditions.This work propels one to think in the direction that $f(T,\phi) $ theories may only allow for weak cosmological singularities, which may make them better placed than some of the other modified gravity based dark energy regimes which allow for more singularities and also those of the stronger types.
\begin{acknowledgements}
The authors would like to thank Sergei Odintsov for very helpful discussions. The research by M.K. was carried out in Southern Federal University with financial support of the Ministry of Science and Higher Education of the Russian Federation (State contract GZ0110/23-10-IF). RCN thanks the CNPq for partial financial support under the project No. 304306/2022-3. This article is based upon work from COST Action CA21136 Addressing observational tensions in cosmology with systematics and fundamental physics (CosmoVerse) supported by COST (European Cooperation in Science and Technology). We would also like to thank the anonymous referee of the manuscript for their insightful comments on the work. 
\end{acknowledgements}

\bibliographystyle{spphys}       
\bibliography{Jcitations.bib}   

\end{document}